\documentclass[sigconf]{acmart}
\usepackage[english]{babel}
\usepackage[utf8]{inputenc}
\usepackage{amsmath}
\usepackage{graphicx}
\usepackage{paralist}
\usepackage{balance}
\usepackage[colorinlistoftodos]{todonotes}
\usepackage[normalem]{ulem}
\useunder{\uline}{\ul}{}
\usepackage[T1]{fontenc}
\newcommand{\cmnt}[1]{\textit{``#1''}}

\usepackage{array,booktabs,arydshln,xcolor}

\usepackage{xcolor,colortbl}
\begin{document}

\copyrightyear{2018} 
\acmYear{2018} 
\setcopyright{acmcopyright}
\acmConference[CPS-SPC '18]{Workshop on Cyber-Physical Systems Security \& Privacy}{October 19, 2018}{Toronto, ON, Canada}
\acmBooktitle{Workshop on Cyber-Physical Systems Security \& Privacy (CPS-SPC '18), October 19, 2018, Toronto, ON, Canada}
\acmPrice{15.00}
\acmDOI{10.1145/3264888.3264897}
\acmISBN{978-1-4503-5992-4/18/10}

\title{Science Hackathons for Cyberphysical System Security Research} 
\subtitle{Putting CPS testbed platforms to good use}

\author{Simon N. Foley}
\orcid{0000-0002-0183-1215}
\affiliation{IMT Atlantique, Lab-STICC, \\ Rennes, France 
}
\author{Fabien Autrel}
\affiliation{IMT Atlantique, Lab-STICC, \\ Rennes, France 
}
\author{Edwin Bourget}
\affiliation{IMT Atlantique, Lab-STICC, \\ Rennes, France 
}
\author{Thomas Cl\'edel}
\affiliation{IMT Atlantique, Lab-STICC, \\ Rennes, France 
}

\author{Stephane Grunenwald}
\affiliation{IMT Atlantique, Lab-STICC, \\ Rennes, France 
}
\author{Jose Rubio Hernan}
\affiliation{T\'el\'ecom SudParis, \\ Evry, France  %SAMOVAR
}
\author{Alexandre Kabil}
\affiliation{IMT Atlantique, Lab-STICC,  \\ Brest, France 
}
\author{Rapha\"el Larsen}
\affiliation{IMT Atlantique, LaTIM, \\ Brest, France 
}
\author{Vivien M. Rooney}
\affiliation{IMT Atlantique, Lab-STICC, \\ Rennes, France 
}
\author{Kristen Vanhulst}
\affiliation{T\'el\'ecom ParisTech, \\ Paris, France 
}

\begin{abstract}
%Cyber-physical system testbeds provide platforms with which to investigate and demonstrate a variety of research questions. 
A challenge is to develop cyber-physical system scenarios that reflect the diversity and complexity of real-life cyber-physical systems in the research questions that they address. 
Time-bounded collaborative events, such as hackathons, jams and sprints, are increasingly used as a means of bringing groups of individuals together, in order to explore challenges and develop solutions. 
This paper describes our experiences, using a \textit{science hackathon} to bring individual researchers together, in order to develop a common use-case implemented on a shared CPS testbed platform that embodies the diversity in their own security research questions.  A qualitative study of the event was conducted, in order to evaluate the success of the process, with a view to improving future similar events.
\end{abstract}

%Simon, how about including something like this: The evaluation demonstrates the value of the process.



\maketitle

%\begin{minipage}{0.4\textwidth}
\hfill \textit{"The hardest single part of building a software system} 

\hfill \textit{is deciding precisely what to build"}

\hfill [Frederick P.~Brooks Jr., ``The Mythical Man-Month'', 1975]
%\end{minipage}

\renewcommand{\shortauthors}{S.N.~Foley et. al.}

\section{Introduction}

Securing Cyber-physical systems (CPS) is non-trivial, requiring techniques and understanding that span a variety of computational and physical components. 
It is often this diversity that is exploited by attackers: 
a siloing of expertise on the part of researchers and/or developers can mean that vulnerabilities are overlooked, introduced or interoperate in ways that are not anticipated~\cite{Pieczul:2017:DSS:3171533.3171539}. 
CPS~testbeds provide platforms that can help understand and investigate research on security techniques, 
and testbeds that reflect the diversity of CPS components have an important role to play~\cite{Antonioli:2017:GIS:3140241.3140253,DBLP:conf/uss/GreenLARHR17}.
However, construction and maintenance of  `real-life' testbeds requires expertise \cite{DBLP:conf/uss/GreenLARHR17,SWaT}
which, along with potentially high capital and recurrent costs, may deter their use in research. 

Sharing research equipment among research groups is one course of action. %\cite{N8report} 
This is more than just sharing infrastructure, it is also about pooling expertise and collaboration.
%Access to research equipment also fosters research ``because it's there'', and ideally precipitates BITNET-effect style benefits \cite{NBERw12812}.
The challenge is \textit{how} to enable this research using shared CPS platforms.

In this paper we consider how time-bounded collaborative events, centred around shared CPS testbed platforms, can be used to enable security research.
These events \cite{Filippova:2017:HMT:3022198.3022658}, such as hackathons, jams and sprints, are increasingly used as a means to bring groups of individuals together, to explore challenges and develop solutions. 
The primary contribution of this paper is a methodology for a \textit{science hackathon} used to develop a common use-case using shared CPS testbed equipment. 
In addition to providing a `real-life' CPS~infrastructure,  the use-case supports diversity in the research questions it addresses. 
This is achieved by drawing together security research from a number of individual projects, including  diagnostics, resilience, visualization and anomaly detection. 
%A qualitative study on the experience of this activity has been used to evaluate the process
% with findings consistent with other work, as well as 
%providing insights  specific to putting CPS testbeds to good use.  

The paper is organized as follows. 
Section~\ref{s:related} discusses some related work on time-bounded collaborative events and 
CPS testbed platforms. 
Section~\ref{s:trans} proposes the use of transverse use-cases on shared CPS testbed platforms as a means of supporting diverse research questions.
Section~\ref{s:hack} proposes the use of a science hackathon as a means to develop transverse use-cases and Section~\ref{s:case} describes a CPS testbed platform that was developed following this process. 
A qualitative study on the experience of this activity, described in Section~\ref{s:eval}, has been used to evaluate the process.
%An evaluation of the process is provided in Section~\ref{s:eval}. 
Section~\ref{s:conclusion} provides discussion and conclusion.

\section{Related work} \label{s:related}

There is a good deal of published literature on time-bounded collaborative events, 
ranging from academic studies of public hackathons \cite{Richterich:2017} 
to accounts by practitioners of their own experiences \cite{Frey:2016:IHO:3011784.3011794,scienceHack}. 
An ongoing theme is the tension between the pressures to produce artefacts versus the idealised notion of a hackathon as a free-wheeling melting pot of creativity, ideas, and skills.
In an ethnographic study of how participants organise themselves and their development practices, Richterich \cite{Richterich:2017} found that the pressures to produce artifacts can skew the activity away from personal learning and technical depth. 
On the other hand, Frey et. al. \cite{Frey:2016:IHO:3011784.3011794} consider how organisational structures can slow innovation and describe their practical experiences of using hackathons as a means to facilitate innovation within the organisation. 

The time-bounded event described in this paper is closest, in sentiment, to the science hackathon \cite{scienceHack} which exposes researchers to new research challenges and helps them to develop their own research ideas within a broader research landscape. Science hackathons are time-bounded collaborative events where a group of researchers come together to identify research challenges and `hack' new lines of research.  
A case in point is the science jam at the ACM Conference on Human Factors in Computing Systems,  with the objective to enable small groups of individuals to research, and develop a research poster within a  two day timeframe.

%Green {\it et al.} \cite{green:testbeddiversity,DBLP:conf/uss/GreenLARHR17} build testbeds with diversity in order to represent the complexity of these systems, and the improvement of security and resilience. \cite{SWaT} propose a Secure Water Treatment Testbed as an example of a real-world ICS, which provides a source of data for experimentation,  underlying infrastructure is generally static. In \cite{Rubio-Hernan:2017}, the authors show how to build a replicable and affordable CPS testbed platform, taking into account the wide ranging complexity of these systems. Similarly, \cite{Siaterlis:EPIC} create an  emulation based system in order to analyze the impact of cyber attack scenarios on the cyber and physical dimensions of the cyber-physical systems. Moreover, to study the cyber-physical systems security we can use simulation based systems, for example, real-time simulation of control processes and physical devices running on the mininet network emulator \cite{miniCPS}. The Acumen testbed \cite{AcumenTestbed:2016} is another example, supporting modeling and simulation of CPS. 

Contemporary hackathons that focus on CPS and IoT development are becoming common, 
although we are not aware of their use in a scientific context for CPS research.
Wagner \cite{Wagner:2014:SCS:2593812.2593819} discusses how the Scrum agile methodology might be adapted in the context of a design-sprint for CPS systems, although it has not been tried in practice. 
Taha et. al. \cite{AcumenTestbed:2016} describe an agile development of an open-source CPS testbed that is 
intended for research and education, and supports simulation and verification of continuous and discrete CPS models. 
Our transverse use-cases go beyond simulation, with an emphasis on working with real-life infrastructure and platforms, concurring with  Green et. al. \cite{DBLP:conf/uss/GreenLARHR17} who highlight that in real-life, CPS infrastructures, such as Industrial Control Systems, comprise a wide range of different equipment and that research testbeds should reflect this diversity.
This emphasis of dealing with real-life infrastructure and equipment for training and research is also seen in the 
development of Capture the Flag style gamification of an ICS testbed platform \cite{Antonioli:2017:GIS:3140241.3140253,iTrust:S3-17}. 
The activity described in this paper is not a conventional competitive hackathon;
as a combination of jam and hackathon, it emphasises hacking CPS platforms as a means to integrate, demonstrate and explore lines of research.

%Contemporary hackathons are becoming relatively competitive events with time-pressures and prize incentives.  This can encourage participants to focus on persuasive presentations and organizing themselves to optimize resources, at the expense of technical depth and personal learning \cite{Warner:2017:HEC:3105726.3106174,Richterich:2017}. This contrasts with the more recent phenomena of the \textit{science hackathon} \cite{scienceHack} where emphasis is placed on learning and collegiality among participants, rather than on the competitive aspects.  Science hackathons are time-bounded collaborative events where a group of researchers come together to identify research challenges and develop new lines of research, that is, to hack research.  A case in point is the ACM Conference on Human Factors in Computing Systems.  The objective of the science jam offered by the conference is to enable small groups of individuals to research, and develop a research poster within a  two day time frame~\cite{chi-jam2018}.

%\section{Putting CPS testbeds to good use: transverse use cases} \label{s:trans}
\section{Transverse use cases} \label{s:trans}

The Cyber CNI Chair %\footnote{\url{https://www.chairecyber-cni.org}}
is an Institut Mines T\'el\'ecom (IMT) industry chair on the cyber-security of Critical National Infrastructure with an emphasis on security of cyber-physical systems. 
A collaboration between three geographically dispersed IMT schools and eight industry partners,  \textit{targeted} industry use-cases form an important part of the research activity of each doctoral/post-doctoral researcher.
With over fifty academic and industry researchers and investigators working together on ten separate, although related, targeted projects, there is a risk that these research efforts become siloed, if not by project, then by school.
%Even with the best intentions for collaboration, researchers, focused on their particular industry targeted use-case, may overlook other efforts and thereby miss  potential synergies with other projects. 
Furthermore, Intellectual Property (IP) constraints surrounding an industry target use-case may make it difficult to share the results within the chair or to the broader scientific  community.  
Foreground research results may become intertwined with commercially sensitive target use-case technology/background IP that an industry partner wishes to protect. 

Building on the research activities across separate projects, \textit{transverse use cases}  are proposed as a means of addressing these risks. 
These use-cases are independent scenarios that are developed around a common cyber-physical system platform, and are intended to build synergy and enable unencumbered demonstration and sharing of research across the chair, as well as to the wider community.
%Cyber-physical systems can have high capital and recurrent costs and, therefore, it is not cost-effective to have separate equipment dedicated to each project and/or geographic site. Developing the transverse use-cases on  common platforms promotes the use of shared equipment as a cost-efficient means  of conducting research on capital intensive systems.  
%A transverse use-case is intended to stimulate interaction among researchers across the chair,  help build an \textit{esprit de corps},  and produce artifacts that can be used  to conduct and show-case ongoing research,  while complementing the existing targeted use-cases.  
The ambition is to build innovative prototypes that span the research projects, while being fail-fast, so as to avoid the sunk-cost-fallacy, as needs be. 
%Based on this overarching goal, the scope of the transverse use-case for each researcher is as follows:
%\begin{itemize}
%    \item A transverse use-case is intended to be a small, rapid prototype that embodies some of the security threats and ideas drawn from a number of research projects.  
%    \item A transverse use-case is not intended as a substitute, or an alternative, to targeted CPS use-cases. 
%    \item The prototype can function to illustrate ongoing research, and as a catalyst for developing synergy between projects and follow-on industrial engagement.  
    % As such the transverse use-case is not necessarily a once-off activity, rather it can evolve and be extended over time.
%    \item The transverse use-case should not distract from day-to-day research work, and is not on the critical path for the individual doctoral/post-doctoral research project.
%    \item The transverse use-case is not intended to entail significant development effort,     rather the emphasis is opportunity-driven and fail-fast.
%    \end{itemize}
We consider transverse use-cases in the spirit of Jackson~\cite{jackson,foley:spw2017}: 
how do we construct an appealing and though-provoking cyber physical system scenario in order to explain and further investigate our research?

\section{Science Hackathons for Security} \label{s:hack}

A science hackathon provides a means to develop transverse use-cases, where researchers come together to explain ongoing research, work on requirements, learn and teach each other about relevant technology and develop rapid prototypes. 
In the following we describe the process that was developed. 
Our approach focussed on collegiality and exploration, as opposed to the development-centric and competitive focus of contemporary hackathons. 
%Researchers benefit by being exposed to new research questions, learning about some technology and exploring how their ideas might fit into a larger research landscape.

%The primary participants in the science hackathon are the doctoral/ post-doctoral researchers, who take ownership of driving the technical details of the hackathon. In this way, these transverse use-cases are intended as a student-run activity. With serendipity in mind, researchers are given as much freedom as is practicable. There is some Principal Investigator involvement, but it is confined to a facilitator role providing guidance and some direction.

\begin{figure}[htb]
\centering
\includegraphics[width=0.45\textwidth]{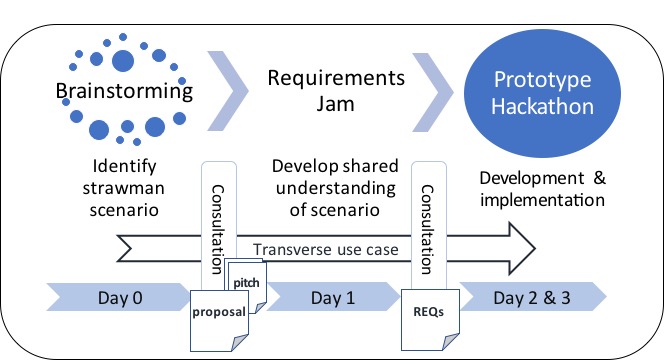}
\caption{The three stages of the science hackathon}\label{f:stages}
\end{figure}

The science hackathon runs across three separate events, as depicted in Figure~\ref{f:stages}.
A 2-hour  brainstorming session develops a strawman proposal based on a potential target CPS platform. 
A one-day requirements jam considers this proposal, and develops a shared understanding of platform requirements, 
which are, in turn, developed and implemented at a 2-day prototyping hackathon.
%The science hackathon is intended as a series of intense collaborative events on security research and development. 
%Studies have shown that developers find it useful to rely on arguments as part of the process of exploring and understanding security requirements and threats~\cite{Weir:2017}. The dialectic argument in this context is a means of encouraging interaction, using counter arguments as a resource, thereby helping to avoid the problematic blind spots associated with group-think or consensus.

\paragraph{Brainstorming}

%The  science hackathon was introduced at the end of an internal research event of the CNI-Chaire where researchers give presentations on their ongoing research activities. 
Following an introduction to the goals of the hackathon, participants spend two hours brainstorming on the available CPS platforms and consider how a platform might form the basis of common scenarios interpreted from their individual research projects. 
The brainstorming session concludes with a short presentation and discussion of a strawman proposal. After the brainstorming session, a short (2 page) proposal document is prepared collaboratively online and circulated for consultation and feedback.

\paragraph{Requirements Jam}

The objective of the one-day requirements jam is to develop shared understanding of the technical requirements for the platform to be developed and how research questions from individual research projects can be cast in this platform.
One week prior to the jam, each participant reviews the strawman proposal and develops a (5 minute) presentation slide with their \textit{initial} answers to three questions: 
\begin{description}
\item[Threat scenario.] 
\textit{What is the threat (related to your research project) that you are focusing on, how do you plan on supporting it in the CPS testbed, and is your approach innovative?} 

\item[Technology challenges.]
\textit{What technical development will be needed on the CPS testbed to implement the  scenario? Are there potential obstacles?
Can your scenario be implemented with minimal (re-)coding of the target (preferable)?}

\item[Research challenge.] \textit{How does this relate to the research questions on your own project? 
Will the platform development leverage/enable your research work (preferable)?} % or is it a new research direction (risky)?}
\end{description}
Each participant uses this to pitch how their work can contribute to the overall scenario. 
Pitches are debated, scenarios are synthesized, revised and/or eliminated, and presented and discussed with partners and investigators at the end of the requirements jam. 
%Participants are encouraged to make a map of everything, documenting requirements. in terms of:  a unified storyboard for the scenario;   how to do it and any specific technical needs for the CPS testbed; their own research activities that it leverages, and  any potential innovations.

Following the requirements jam, a requirements document detailing the CPS threat scenario is prepared and circulated for feedback.
%This living document is a synthesis of the individual pitches that evolved during the course of the jam. 
This is a synthesis of the individual threats identified by participants and cast as a unified scenario for the testbed. 
Additionally, the document identifies required CPS technologies and development, and
the research (project) challenges it is envisaged that the platform will support. %that are envisaged supported by the platform. 
The document prioritizes  tasks to be carried out, whereby tasks with lower priority can be dropped during prototyping if necessary due to time constraints or feasibility concerns

%With its mix of hardware, software and infrastructure components, CPS platform development is time-consuming and requires a high degree of expertise. 
%There is a risk that the hacking/development of the CPS platform can become a means in itself, to the detriment of demonstrating the project-related research challenges in the context of the platform. With this in mind, the requirements document prioritizes the  tasks to be carried out, whereby tasks with lower priority can be dropped during prototyping if necessary due to time constraints or feasibility concerns.
%An overarching principle is that requirements should be such that a prototype can be expected to be constructed during the 2-day implementation hackathon. 

\paragraph{Prototyping hackathon}

The requirements document provides a tentative roadmap for the two day prototyping hackathon.  
Impromptu groups formed around the requirements, solving problems as they arose. 
Studies have shown that anticipation of having to prepare and give presentations can interfere with the technical focus of participants in a hackathon \cite{Richterich:2017}.
Therefore, since the primary objective was to produce technical artifacts for the transverse use-case,  a programme of presentations was avoided and the two-days concluded with a short and informal debriefing session. 
Documentation and presentations of the results were left for after the hackathon.

%\redremark{}{I was pretty hands-off here, though other than adding something about the physical arrangements I don't think there's much else to say here. Is there anything else in particular that you think would be useful to add here to 'describe' your process of arranging yourselves? }

\section{Case study} \label{s:case}

A science hackathon was conducted during the first half of 2018, following the three stages described in the previous section. 
The brainstorming session (February) identified a Fischertechnik-based CPS testbed as the basis for the strawman scenario.
Over the course of the requirements jam (May) and the prototype hackathon (June), the testbed was adapted to be consistent with project research questions. 
Seven doctoral researchers from seven different projects participated, along with one research engineer who provided domain expertise on the testbed. One Principal Investigator facilitated the process and an Applied Psychologist conducted a study of the experiences of the participants. 
This section gives an overview of the transverse use-case that was developed. 

%\redremark{}{The following subsections are intended to loosely follow the structure of the requirements document generated by the requirements jam. That is, a storyboard for the the synthesized threat scenario; the CPS technology requirements for supporting the threat scenario, and the research challenges to be supported by the platform/threat scenario }

\subsection{A Fischertechnik-based CPS testbed}
\label{sec:testbed}

%\begin{figure}[htb]
%\centering
%\includegraphics[width=0.45\textwidth]{CPSpicture.jpg}
%\caption{Fischertechnik CPS testbed}
%\end{figure}

\begin{figure}[htb]

\centering
\includegraphics[width=0.4\textwidth]{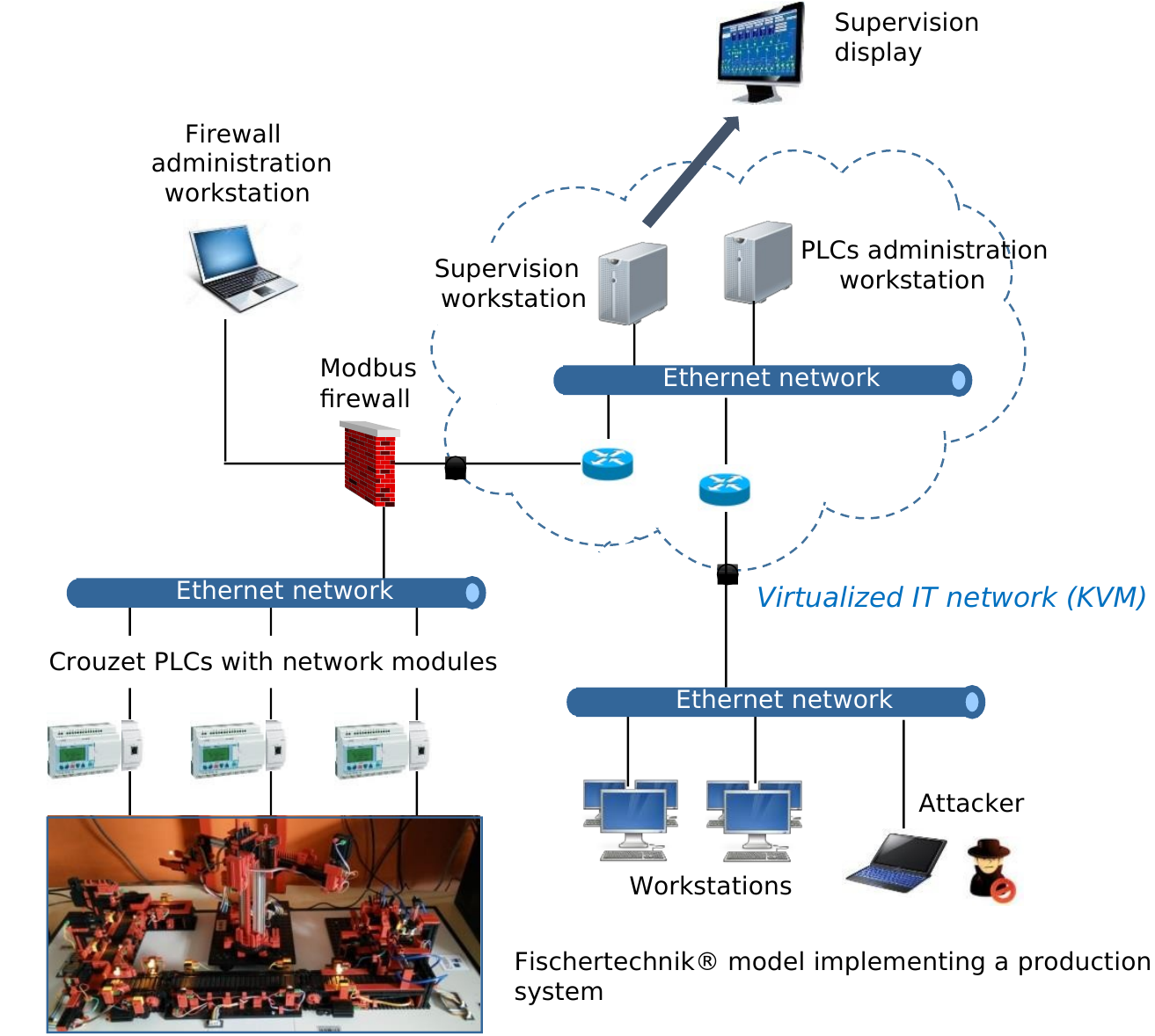}

\caption{Fischertechnik CPS architecture} \label{f:architecture}
\end{figure}

The testbed used for the hackathon reproduces, at a small scale, how modern industrial control systems are connected to a organization's IT network, 
as depicted in Figure~\ref{f:architecture}. 
%\begin{itemize}
    %\item 
    A virtualized IT network with two VM border routers isolate the network from the Internet and the production network, and 
    two VMs provide a supervision system and an administration system, to control and monitor the PLCs which drive the production line, respectively.
    %\item 
    A production network connects production line PLCs.
    %\item 
     Network-enabled Crouzet PLCs support modbus requests from virtualised supervision/administration workstations.
    %\item 
    An industrial firewall supporting modbus and S7, is configured to enable modbus communications between the PLCs and the supervision and administration workstations.
    %\item 
    Custom grafcet programs provide automation for the Fischertechnik platform, a machining production system with conveyor belts, DC motors, a pneumatic press, a drill, a mill, a robotic arm which manipulates the processed parts and assorted mechanical and optical sensors. 
    Parts to be processed move from one machining tool to the next, using conveyor belts, pistons and motors. The processed parts are plastic cylinders that are moved by the robotic arm from the end of the line to the beginning so that the process run continuously, as required, without  human intervention.
%\end{itemize}

\subsection{Threat scenario}
\label{sec:threatScenario}

The requirements jam storyboarded a threat scenario that framed  research questions from different projects. 
It comprised an external attacker exploiting weaknesses in the network perimeter defences and/or an internal attacker with direct access to IT or OT components. 
The attacker has two objectives: 
(1) stop a conveyor-belt or release the clamp to halt the production line (easily noticed), 
or (2) change milling/drilling times in order to reduce  finished product quality (more subtle).
Both objectives require the attacker to send Modbus packets to at least one Crouzet PLC situated in the OT part of the organization. 
As only a specialized insider could physically access OT, other attackers need to take control at the OT administration workstation. 
Two attack vectors were considered: 
(1) exploit a vulnerability of the workstation or obtain administrator credentials via a dictionary or password-guessing attack, 
and (2) disable (physically or remotely) the workstation, forcing administrators to use a less secure secondary rescue workstation.
%(non updated, with a weaker configuration or with weak credentials).

\subsection{Technology challenges}

%\SF{jam identified the potential challenges, hackathon tried them out/ engineered them}

In implementing the threat scenario, testbed changes were necessary in order to enable researchers deploy their tools and investigate their research questions. 
The need for these changes were identified during the requirements jam, and implemented during the hackathon, along with other unanticipated changes, as they emerged.  
Further switches were added to the production network, supporting additional hosts, an Intrusion Detection System, network traffic sniffers and hosts to inject network traffic. 
%Regarding the virtualized IT network, virtual machines were added. 
The PLC grafcets were modified to expose some of their state variables in modbus registers.
%A threat scenario was necessary that could encompass all of the research questions/projects on the testbed platform.
%However, the threat scenario in itself was not deemed sufficient to illustrate all of the research projects the platform was to be used for. 
The Fischertechnik platform was extended to incorporate an interconnection with a second virtualized industrial system, as part of a larger industrial process. 
%, a  fictitious cream enrichment plant that extracts cream from raw milk. 
%The existing testbed was extended to support this new scenario, by including an additional industrial systems. 
%Therefore, we came up with another scenario where we would have two industrial systems : a cream enrichment plant producing cream from milk,
In this scenario, the Fischertechnik platform represents a manufacturing plant that builds a replacement component upon its failure in the virtualized industrial system.  
The state of the virtualized industrial system is modelled in terms of a collection of PLC/Modbus registers (using the EasyModbus library), that the SCADA system monitors and controls the Crouzet PLCs/Fischertechnik platform as appropriate. 
Virtualization is a good compromise, in term of resources expended during the hackathon, and facilitates making, in a realistic way, more complex scenarios for the research projects that deal with supervision and system modeling.

This setup of the testbed was used to support the threat scenario described in section \ref{sec:threatScenario}. 
The fischertechnik process is triggered on failure of a component in the virtualized ICS. 
This could represent normal wear and tear, or an accident.  
However, it could also be a result of an attacker interfering with the production processes, forcing early failure.
Such complex situations with interleaved safety and security call for more advanced diagnosis methods and are opportunities to illustrate the models developed in the research challenges the platform was designed for. 
A secondary workstation providing backup administration, configured with an old 
version of windows with known vulnerabilities, along with weak credentials, provides the (second) attack vector.  

%Indeed, the failure of the centrifuge could be purely accidental but then an attacker could alter the production of the replacement component. Or the attacker could voluntarily increase the rotation speed of the centrifuge to increase its failure rate but then the production of the replacement component could encounter a problem. 

%With this in mind, the network has been modified to give a distinct IP address to this new virtual machine and IP filtering rules of the firewall configuration have been changed to allow only the rescue station and the original administration workstation to communicate with the OT part of the organization. 
%Moreover, both attack scenarios need the attacker to be able to take control or at least to disable the administration workstation. But this virtual machine is originally running on Windows 10 and we decided to replace it by a version of Windows 7 with known vulnerabilities. The rescue workstation also runs on this same version but weak credentials have been set up.

% Add IDS part

\subsection{Research challenges supported by testbed}

%\redremark{All}{a series of sketches of how aspects of the research challenegs in your individual research projects can be described in terms of the adapted testbed.  Around a paragraph of material per project, preferably synthesized together.

%see alternative text below
%SS \RL{} Knowing the source  and the path of information and when pieces of information had been fraudulently or undesirably modified can be referred respectively as traceability and integrity of information. In industrial systems which are often legacy systems, this represents a serious challenge in addition to the fact that the components of these systems are often constrained by their power consumption. That is why we need passive solutions for the information analysis, before as well as after the attacks/failures. 
%Constructing together an end-to-end attack on the testbed helped us to consider some aspects of security that some of us did not think of at first while others did and vice versa.  The product of the hackathon can be used to illustrate the issues of integrity and traceability of information along with testing passive solutions.

A primary objective of the science hackathon was to develop a CPS platform in which research questions from diverse projects could be demonstrated and explored. 
%The testbed is used to evaluate diagnosis models that provide explanations when incidents occur. 
By providing distributions for the probability of component attacks and failures, the model proposed by Bourget et. al. \cite{bourget:2018} can be used to provide real-time diagnosis of security and safety in the testbed.
Testbed snapshots of the state of the PLCs, along with alert events from the IDS enable diagnosis of the probability of attack/failure as the threat scenario (Section~\ref{sec:threatScenario}) evolves.
This can be used to identify the origin of an incident that generates several alerts, compute probabilities of occurrence of future events or evaluate the likelihood of any given event. Such information can be used to decide which is the appropriate response to the incident.
These PLC state snapshots and IDS events are also used in another project investigating anomaly detection. %on the data from sensors, actuators, and other components, and their correlation with the system's state/configuration. 

%SS For instance, the scenario described in section \ref{sec:threatScenario}, purely security, can be embellished with a components failures and their consequences in order to drive a safety assessment of the platform. It can be modeled in a diagnosis model such as the one described by Bourget \textit{et al.} [REF]. Once the event model has been fully described, it can be plugged in the SCADA and the IDS in order to instantiate events and have a live photography of the state of the system at any moment. 

%SS - pretty much said already. The hackathon was a real opportunity to identify and implement attack scenarios on the system. These scenarios were defined according to several levels of skills and resources provided by the participants. 
%KV I clarified the following points : can you provide more context for this? what is the risk model you refer to? Is it important that one phase is easier than the other? Perhaps focus on what it is you are describing here.
%\KV{}
During the implementation, a counter-intuitive observation of the risk model was recognized. The chosen approach was split into two phases. The first phase consists of recognizing the network as well as the Modbus registers. The second phase consists of finding the effective attacks and executing them. Intuitively, the first phase seems to be easier than the second one. In contrast, the recognition phase was much more time consuming than the other phase (according to our implementation). This kind of information is useful especially in the context of risk analysis. Indeed, it allows defining
%to define 
a likelihood of success for each attack (which corresponds to the risk model) in order to carry out the evaluation of the overall risk. In fact, this result highlights the importance of the sensitivity analysis on risk assessment. Moreover, such analysis should take into consideration assumptions on the risk model.

The testbed is a complex arrangement of hardware and software. 
In its design, various mechanisms help enhance its resilience to attack and failure.  
For example, network virtualization helps provide defense in depth, while sensor redundancy helps provide fault-tolerance. 
We use \cite{cledel:2018} to model deployments of these mechanisms in order to determine, and compare, their collective effectiveness at providing resilience in the testbed. 
For example, the rescue station is configured with weak credentials that allow an attacker to make a dictionary attack. If these credentials are replaced by stronger ones, how is the overall resilience affected?
%Moreover, once the CPS is described and an attack scenario is determined, it can be planned to make changes to the CPS in order to remedy the attack. If several solutions are considered, a resilience evaluation of the system and of the different solutions can determined which one is the best to improve the system and how better a solution is compared to the others. Besides, a resilience evaluation can also prevent unexpected side effects of a remedy by evaluating if the system will still fulfill its functions, critical or secondary, if the remedy is applied and installed.

%\AK{useful to refer to your security cyberCOP paper here?} 
Immersive 3D visualization techniques can provide insights into the security and safety of the system that go beyond the more conventional  mechanisms and techniques \cite{kabil:2018}. 
%By having testbed fully described by scenarios and models, we are able to simulate its 
We are using a 3D virtual environment to simulate testbed configuration, behavior and security events. 
This kind of simulation can help us to better understand the compatibility of our different models (diagnosis, source and path information, resilience) and to visualize the threat scenarios in a more graphical and interactive way.

\section{Experience of the hackathon} \label{s:eval} % lessons learnt

\begin{quotation} \it
``If someone told us, do that,
it wouldn't be a hackathon.''
\hfill \em [Interview extract]
\end{quotation}

%\VR{To finish introduction}

The evaluation aimed to understand team member experience of the science hackathon, with a view to applying insights gained to improve the process in future events. 
% For the hackathon, the team met together on three occasions.  The first facilitated a half day of work, the second allowed for one full day, and the third and final meeting took place over a two day session. %Methodologically, we wanted to reflect the nature of the hackathon, thus 
It was decided to conduct semi-structured interviews with individual team members during the second day of the prototype hackathon while the team were still working together.  
%The goal of the individual interviews with team members was to gain insights from all participants.  
An interview schedule was developed in conjunction with the organiser of the event. %and in light of the literature review.  
Thematic Analysis \cite{richie:2003} was applied to the audio recordings. 
Individual recordings were summarised as transcribed text, emerging themes were identified, and structured the analysis.  
Transcribed material was anonymised and deidentified, and short verbatim extracts used to illustrate the analysis \cite{oconnell:1995}.  

An ethical self evaluation was completed.  The process of consent was initiated with participants via email, providing team members with information on the proposed evaluation. 
This was followed by a second email, with an information sheet on method and consent form, for consideration. The proposed data collection was explained verbally to participants on Day~1 of Stage~3; as was informed consent, such as anonymity and that there was no onus to participate.  This was an opportunity for questions from the team; questions and comments were also invited prior to, and following, individual interviews.
A consent form was signed prior to each interview by both researcher and participant, and each retains a copy.  The forms retained by the researcher are held securely.  
%\cite{guillemin:2009} 
% For addition to the bib: \cite{guillemin:2009} is: M.Guillemin and K.Heggen.  Rapport and respect: negotiating ethical relations between researcher and participant.  Medicine, Health Care and Philosophy, 12(3):291--299, August, 2009.
The interview data comprises a total of 141 minutes, with interview duration ranging from 15 to 30 minutes, with an average of 20 minutes.  
Themes that emerged from the analysis are discussed below.

%\subsection{The three stages of the science hackathon}
\subsection{A process for hacking scientific research}

\paragraph{Expectations} 
The nature of science hackathon, as being a scientific collaboration, had been clear in the information circulated at the outset, and this was reflected in the expectations of the participants, as expressed by the comment: \cmnt{I asked colleagues, and they said a hackathon was when you program all day, but the scientific hackathon was explained by Simon, so I understood that it was not about creating a product, but to work with and talk to other people}.

%However, while the goal of working with and talking to people was clear, there was, nevertheless, some underlying concern about the expectations of those outside the hackathon team.  This concern is about the production of a fully integrated end-to-end result.  One participant said, for instance, that he understood that what is produced for Simon is acceptable, but that may not be case for everyone. For example, that others \cmnt{may expect something concrete, something that can be shown, and I am afraid that we will not have what they expect, even if we have something, we will not have enough for them}.

\paragraph{Requirements Jam}%{Preparing for the prototyping hackathon}

The preparatory work required for the jam was reported as not being specific enough, in terms of substance and purpose.  For instance, some thought that the questions for the pitch were too broad, and were unsure if their response was as expected.  However, when the pitch slides from individual participants were shown at the second meeting, the usefulness of the requirements jam was clear.  The preparatory work was seen as an opportunity to clarify individual participation, one participant comments that 
\cmnt{%Simon told us we only had to work on the hackathon when we were here, but 
......we had to prepare this part and to write it down, we didn't use it much here, but writing it down made it clearer to us, in our heads}.  This was in terms of illustrating the different research perspectives on the same issues, and the identification of shared goals. %, namely, the reason for them being there together.  
In another participant's words, \cmnt{we can work together in a holistic way, individuals can do parts, then blend}.  

\paragraph{The extended time frame}  For the team, being together over a longer time frame than usual facilitated cohesion, both professionally and socially.  This is illustrated by the remark: \cmnt{Usually we have one day meetings, and you cannot talk with people a lot about the work, and the hackathon gives the chance, there is more time to talk to people about their work, about my work, and also not in a professional way, so that is good}.

%\subsection{Broadening scope of individual research}
%\subsection{Linking disparate research strands}
\subsection{Transverse use-case: avoiding research silos}

\paragraph{Linking research strands.}%{Collaboration with others}
Participants talked about the collaborative aspect of the hackathon as being interesting, creative, enjoyable and useful.  Working with, and learning from, others who have different perspectives and expertise in Cyber Security was engaging.  Furthermore, interaction sparked creativity as illustrated by the following comment:  \cmnt{It gives me ideas, working with everybody, and we produce stuff that we can use for their work, for my work, so I am very pleased}. 
%The usefulness of the collaborative aspect of the hackathon for contextualising 
A benefit for research in the Chair flowing from the opportunity to link disparate research strands was reported, as illustrated by the following: \cmnt{Interesting to find out about the diverse knowledge that each PhD student has, for example, good at hacking, good at graphing {\em [PLC grafcets]}.  We have different knowledge, and put together, can create a different point of view, and explain better what they are doing in the Chair.}  As the above illustrates, the process of collaboration and making links was something that the individual researchers appreciated and valued.  %In general, this process was conceptualised as being separate from any output that might be produced: the participants worked well together.  The experience was also enjoyed by participants, as the following comment illustrates:  \cmnt{We all know each other, so working on the hackathon team is fun.}

\paragraph{Positioning individual research projects}
%\paragraph{Relationship with research projects}
While one goal of the hackathon was to foster collaboration %among the researchers 
in the Chair, additional practical and conceptual benefits were identified for participants' individual work.  For instance, being able to develop a practical scenario for their own research, exemplified by the following comments: \cmnt{What I wanted was to get an opportunity for a realistic scenario {\em [for my research]}, and we did it yesterday, so that worked for me} and also \cmnt{the platform is the best way to show people your work and aims.}  The conceptual benefit for individual PhD research is the reciprocal input %from team members 
into each others' research, helping to clarify PhD research, %for individuals
as illustrated by this comment: \cmnt{Being in the same room for two days and chatting, helped me to think about what I will put in my PhD.}

%\subsection{Transverse use-case: avoiding research silos}

\paragraph{The function of the transverse use-case} 
Participants saw the platform %, and what they were working on in the hackathon, 
as being a useful resource to demonstrate and explain their own, and others', research, and the possibilities it entails. 
%, as well as being a means of fostering collaborative work in the future.  Pedagogically, the transverse use-case was a success, as participants learned how the platform could be useful directly and indirectly for research.  %Participants were aware of how the CPS platform could, and could not, be used to illustrate/explain their own work, as well as the work of the other participants.  
For some, the potential use of the CPS platform is in communicating the meaning of an attack, and its significance.  
For instance, being able to demonstrate an attack to non-technical people is useful, as the following illustrates: 
\cmnt{It's very interesting to use {\em [the CPS platform]}, because we can see the attack, we were able to run some stuff on the platform, we started an attack and the platform started moving the wrong way, interesting stuff, and for people that can't understand the technical stuff, seeing the platform moving the wrong way, it's very understandable, and you can say, that is the attack.}

%Demonstrating the significance of a cyber attack is a potential role identified for the CPS platform, illustrated by the following comment: \cmnt{It's a show, just shows how an attack can be devastating, powerful, and this is the importance of the attack in the show.}

\subsection{CPS platform as a communication resource}

%Participants were enthusiastic about the potential of the %hackathon and  platform for their PhD research, as well as for supporting new collaborative projects %in the longer term, within the Chair. 

%The participants were aware 
Conscious of their role as researchers working with industrial partners in the Chair, %.  As such, they 
the participants envisage the platform as a means of facilitating communication to those industrial partners.  For example, one participant talked about how %what they did during the hackathon, namely 
%they had previously failed to find ways to make convincing prototypes, %they had tried previously without success, 
they had not previously been successful making convincing prototypes, 
while at the hackathon, they succeeded with one simple attack scenario.  
Also, being able to convey to industrial partners what is possible and feasible from a research point of view can be challenging for participants.  
%An example is the feasibility of conducting proposed research within the time frame that is available for a project.  
An example is communicating both the diversity and the broad scope of Cyber Security and, therefore, how the expertise of an individual is limited within the breadth of the domain.    

Using the platform, and ideas from the hackathon, for communication with industrial partners is envisaged as being a reciprocal process.  As such, having feedback %from the industrial partners 
is important to the participants, as the following explains: 
\cmnt{Having an industrial partner involved would be good, especially for feedback.}
The hackathon is seen as a way of linking research to the real world.  
The CPS platform, and how it was used in the hackathon, are seen as a means of improving communication between researchers and industrial partners.

%\subsection{The future}

%Participants viewed the hackathon as a way of thinking about future research and collaborations, and as an important way of mixing their own research with that of others in the Chair.  Indeed, one participant explained that: 
%\cmnt{I have exchanged {\em [research]} ideas with one of the others already.}  Future hackathons were proposed as a way of linking the current research projects in the Chair, with a view to creating follow-on and new scientific research projects. A further proposed use of the scientific hackathon process is for explaining the research, or  potentially training people in Cyber Security.

\subsection{Autonomous collaboration}

When asked a general question about what was the best part of the hackathon, participants responses focussed on the enjoyment of working in a team, and being challenged, while not under pressure to produce a integrated end-to-end artifact, as such.  As PhD researchers generally working on their own within a supervisory framework, the hackathon provided an opportunity for them to be able to interact with peers, to be curious about others' work, and to learn about it, even if not understanding everything.  The following comment exemplifies the sense of pleasure of the team work:

\cmnt{I enjoyed the challenge of the hackathon, as Simon presented it, having to mix all the work, very interesting for me, the discussions were very interesting, I started the hackathon with great enthusiasm because of the idea of the scientific hackathon, I wanted to do something, talk with others, work in a group, because in a PhD you don't work with colleagues, so it was very interesting to work on the hackathon, new ideas, why not?}

%And another comment highlights the short term intensity of the hackathon team work, in contrast to the usual more solitary experience of the participants:

%\cmnt{It's quite fast, not long, so we work all day, it's more intense, most of the time we work alone, {\em [in the hackathon]} we talk a lot, exchange ideas, everybody, everybody gives an idea, so everybody was involved in the work.}

The freedom that the unstructured nature of the prototyping hackathon facilitated was a very positive aspect of the experience.  The lack of pressure to produce a specific deliverable meant that the team felt able to try ideas and discard them quickly.  The following extract from one interview explains this:  

\cmnt{Thought it might be managed, and it wasn't, which is good, given an area to play, we felt free to do the best we could do, we didn't have a specific aim to reach, so whether it's a success or a small success, we don't feel much pressure about it, we feel free to try to things, if they work, good, if they don't work, try something else, this is good.}

%The cohesion among the researchers that the hackathon engendered was important, and the following comment explains this well, using the concept of drawing the individual researchers together, physically and conceptually:

%\cmnt{Being all together around the table, on one subject, and exchanging, [....] it's a real collaboration, everyone with their ideas, that's great, the idea of hackathon around the PhDs is great, especially because the subjects are very close to one another, so we can have a centre of gravity.}

% Lessons learnt on testbed development \cite{DBLP:conf/uss/GreenLARHR17}

\section{Conclusion} \label{s:conclusion}

This paper described how  a science hackathon, centred around testbed platform development, can be used to investigate security scenarios that reflect the diversity and complexity of real-life cyber-physical systems in the research questions that they address. 
Unlike a conventional hackathon/Capture the Flag event, the activity was coordinated across time-bounded collaborative events: brainstorming, requirements jam and prototype hackathon. 
Driven primarily by the PhD students, the autonomous and non-competitive nature of the event was beneficial, an observation consistent with other studies of time-bounded collaborative events \cite{Richterich:2017,Frey:2016:IHO:3011784.3011794}. 
%With a further objective of fostering collaboration and an esprit du corps, the science hackathon was successful, as evident in the interest and enjoyment expressed by participants in learning about the work of other team members.  
Ameliorating the risk of siloing research was also addressed effectively, as the transverse use case provides a broader context in which to understand and relate individual research challenges. 

The testbed platform, and what was developed in the hackathon, was conceptualised as a means of communicating with others about research, such as industrial partners, and those with less technical domain expertise.  As such, its role in improving communication is envisaged as being bidirectional, with the suggestion from participants that it could be useful as a way, not alone to explain their research, but also as a way for industrial partners to provide feedback on that research.  This finding is a good use of the testbed, one of the main goals of the event.

Remedying the uncertainty expressed by participants around their expectations for the hackathon highlights an interesting area of tension. 
With a conventional hackathon, expectations are generally well understood: attack or defend.
However,  expectations are less certain in a science hackathon, as was evident in interviews. 
While introducing more organization may help provide clarity, it can work against innovation  \cite{Frey:2016:IHO:3011784.3011794}. 
This motivated the requirements jam as a means to help set expectations, while encouraging autonomous collaboration. 
While being broad in its questions to the extent that individuals were initially somewhat unsure about how to respond, the jam did, 
nevertheless, provide clarity to individuals on the potential for them to contribute.
%, as well as allowing them to see how responses to the same questions differed.  
%A level of uncertainty about what was expected, and what might emerge, from the process was evident. 
%This is coherent with the decision to adopt a light approach to managing the event, as a way of facilitating both serendipity and the opportunity for team independence.  
As such, this was a valuable learning experience for participants, providing an opportunity to develop a sense of their own research in relation to that of others; 
how collaboration could work as a process, as well as for future research.  
Related to this is the uncertainty around the degree of the expected outcome, specifically, whether the conceptual collaborative work was required to culminate in a functioning artifact in order for the hackathon to be deemed a success.  
While providing additional guidance and direction, other than the level of advice that was given, would have removed some of the uncertainty experienced, it may have done so at the expense of the independent learning gained in this collaborative context.  
How best to achieve the best balance in this area of tension is an area for future research.

\paragraph{Acknowledgements} 
%Thanks to Heiko Mantel (Darmstadt) for sharing his own experiences establishing a series of cross-project reference scenarios for the DFG Priority Programme~RS$^3$.
This work was supported the Cyber CNI Chair of Institute Mines-T\'el\'ecom which is held by IMT Atlantique and supported by Airbus Defence and Space, Amossys, BNP Parisbas, EDF, Orange, La Poste, Nokia, Soci\'et\'e G\'en\'erale and the Regional Council of Brittany; it has been acknowledged by the French Centre of Excellence in Cybersecurity.

\balance
\bibliographystyle{ACM-Reference-Format}
%\bibliography{bib}

\end{document}